# Real-time decision-making for autonomous vehicles under faults

Xin Tao
*Integrated Transport Research Lab (ITRL)*
*KTH Royal Institute of Technology*
*Stockholm, Sweden*
taoxin@kth.se

Zhao Yuan
*Electrical Power Systems Laboratory (EPS-Lab)*
*University of Iceland*
*Reykjavik, Iceland*
zhaoyuan@hi.is

*Abstract*—This paper addresses the challenges of decision-making for autonomous vehicles under faults during a transport mission. A real-time decision-making problem of vehicle routing planning considering maintenance management is formulated as an optimization problem. The goal is to minimize the total time to finish the transport mission by selecting the optimal workshop to conduct the maintenance and the corresponding routes. Two methods are proposed to solve the optimization problem based on two methods of fundamental solutions: (1) Mixed Integer Programming; (2) Dijkstra's algorithm. We adapt these methods to solve the optimization problem and consider improving the computation efficiency. Numerical studies of test cases of highway and urban scenarios are presented to demonstrate the proposed methods, which show the feasibility and high computational efficiency of both methods.

*Keywords*-Decision-making; autonomous vehicles; fault; route planning; maintenance management

## Nomenclature

*Indexes and Sets*

| | |
|---|---|
| $i, n$ | Node indexes. |
| $j$ | Link indexes. |
| $k$ | Route indexes. |
| $\mathcal{A}$ | Adjacency matrix of links and nodes. |
| $\mathcal{L}$ | Link set. |
| $\mathcal{N}$ | Node set. |
| $\mathcal{N}^w$ | Workshop node set. |
| $\mathcal{N}^t$ | Traffic node set. |
| $\mathcal{R}$ | Route set. |
| $\mathcal{W}$ | Weight matrix of the road network graph. |

*Parameters*

| | |
|---|---|
| $D_j$ | The length of link $j$. |
| $V_{k,j}$ | The average driving speed of a vehicle on link $j$ in route $k$. |
| $T_i^{st}$ | Time of scheduling a tow truck by workshop $i$. |
| $T_i^m$ | Maintenance time of the breakdown vehicle in workshop $i$. |

*Variables*

| | |
|---|---|
| $x_i$ | Binary variable indicating whether workshop $i$ is selected. |
| $y_{k,j}$ | Binary variable indicating whether link $j$ is selected in route $k$. |
| $T_k$ | Driving time of route $k$. |
| $T_{k,i}^{min}$ | Shortest driving time of route $k$ by selecting workshop $i$. |

*Notations*

| | |
|---|---|
| $T$ | Total time to deliver. |
| $T^{st}$ | Time of scheduling a tow truck. |
| $T^m$ | Time of maintenance in the workshop. |

## I. Introduction

With the rapid development of sensing technology, big data, and artificial intelligence, much effort has been put into developing highly autonomous vehicles in academia and industry. As a core enabler of autonomous vehicles, the real-time decision-making capability is essential and challenging. For autonomous vehicles, decision-making, as a broad concept itself, is mainly focused on navigation in a realistic environment safely and reliably [1]. As for decision-making of vehicles under faults, research efforts are on vehicle-level safety, such as developing collision avoidance systems [2], while system-level concerns are not sufficiently addressed.

The system-level concerns of decision-making of autonomous vehicles are multi-folded. With the development of technologies like vehicle remote health monitoring [3], Vehicle-to-everything (V2X) [4] and traffic monitoring [5], vehicles will have access to much more data and flexible resources. The vast amounts of data and information make it difficult for human beings to process and integrate in real-time and make optimal decisions. For autonomous vehicles at SAE automation level 4 or 5 [6], there will be no driver on board involved in the decision-making process when a fault occurs. As a result, it is essential to automate the real-time decision-making process of autonomous vehicles under faults.

Furthermore, instead of focusing on the safety and reliability of vehicle operation, integration of transport infor-mation and maintenance resources can contribute to making better decisions for vehicles under fault. Maintenance planning, as a typical decision-making problem, is usually considered on a large scale or in a long-term horizon [7]. In [8], the maintenance planning problem of autonomous vehicles during the mission was addressed, which, however, considered a simply transport and road network scenario. During a transport mission of autonomous vehicles, unexpected maintenance demands can increase due to increasing operation duration and need to be considered in a transport environment.

In this paper, we analyze the mode shift of maintenance management from human-driven vehicles to autonomous vehicles. Then we identify the challenges of real-time decision-making of autonomous vehicles considering maintenance management. Then the decision-making problem of an autonomous vehicle breaking down during a transport mission is formulated as a route planning problem of selecting the optimal routes and the optimal workshop. There are plenty of methods to solve complex route planning problems [9] [10]. In this paper, we adopt two efficient methods, including mixed integer programming [11] and Dijkstra's algorithm [12] to the decision-making problem and compare them. The contributions of this paper include:
1) Analysis of the mode shift of maintenance management and the challenges of decision-making of autonomous vehicles considering maintenance management;
2) Formulation of a decision-making problem considering maintenance management as a vehicle route planning problem;
3) Proposing two approaches to model and solve the route planning problem based on mixed integer optimization and Dijkstra's algorithm;
4) Comparison and analysis of the two proposed approaches regarding computing efficiency and extendability.

The remainder of the paper is structured as follows. In Section II, the mode shift analysis and challenge identification are provided. In Section III, a decision-making problem is formulated and two methods are proposed to solve the problem. In Section IV, numerical studies and discussion are presented. Conclusion and future work are discussed in Section V.

## II. MODE SHIFT OF MAINTENANCE MANAGEMENT OF AUTONOMOUS VEHICLES

When a human-driven vehicle becomes autonomous, the mode of maintenance management of the vehicle also changes, which brings new challenges of real-time decision-making of vehicles with unexpected maintenance demands during a transport mission.

A typical operation mode of human-driven trucks is shown in Figure 1 [13]. As shown in the figure, the 'planned operating time' includes four periods, the 'operating time', the 'waste' time, the 'unplanned maintenance' time, and the 'planned maintenance' time. The former two periods, the 'operating time' and the 'waste' time constitute the uptime. The latter two periods, the unplanned maintenance and the planned maintenance constitute the downtime. In the 'not intended for use' time, the truck is at a standstill, which may be due to a range of reasons such as an accident, vehicle malfunctions, or driver illness [14].

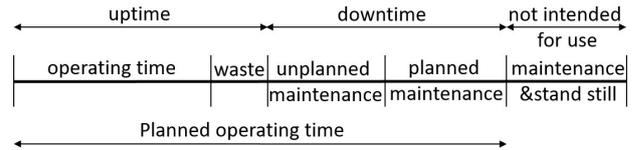

Figure 1. Operation model of human-driven trucks.

When a vehicle becomes autonomous, several mode shifts regarding maintenance management emerges:
1) There is no human driver to check the vehicle to identify abnormalities of the vehicle;
2) There is no human driver to make a maintenance decision and communicate with others when the vehicle encounters faults during a mission;
3) Since an autonomous vehicle is expected to work much more intensely, more unplanned maintenance is likely to be required;
4) There is no human driver to take a short or long rest, or shift in the planned operating time, which reduces the 'waste' time. The standstill time caused by the driver's unavailability no longer exists.

Due to these mode shifts, there is an increasing need to develop automated decision-making methods for autonomous vehicles under faults during a transport mission. The contents of such decision-making may include arranging maintenance activities, vehicle rerouting, and mission redistribution, and so on.

## III. VEHICLE ROUTING PLANNING CONSIDERING MAINTENANCE MANAGEMENT

### A. Problem formulation

In this section, a decision-making problem of vehicle rerouting considering maintenance management is formulated as an optimization problem. We consider that an autonomous vehicle breaks down during a transport mission. In this situation, assume that a workshop will schedule a tow truck to tow the vehicle back to the workshop for maintenance, after which the vehicle will continue and finish the transport mission. In total, five events cost noticeable time before delivery in time order, namely:
1) scheduling the tow truck;
2) tow truck driving to the breakdown site;
3) towing the breakdown vehicle to the workshop;

4) maintenance in the workshop;
5) vehicle finishing the transport mission.

Among them, events 2), 3) and 5) correspond to the driving time of different routes, i.e. the route from the workshop to the breakdown site, the route from the breakdown site to the workshop, and the route from the workshop to the customer. We denote the route set as $\mathcal{R} = \{1, 2, 3\}$, where route 1, 2, and 3 corresponds to the aforementioned three routes. The ending location of route 1 and the starting location of route 2 are both the breakdown site, while the starting location of route 1 and route 3 and the ending location of route 2 are the selected workshop. However, the optimal workshop and the optimal paths to drive in these routes are unknown. We aim to find the workshop that schedules the tow truck so that the time it takes for the vehicle to finish the transport mission is minimal. The expression of the total time to deliver is:

$$T = T^{st} + T^m + \sum_{k \in \mathcal{R}} T_k \quad (1)$$

where $T^{st}$ and $T^m$ are the time of scheduling a tow truck and the time of maintenance respectively, and $T_k$ is the driving time of route $k$.

To find the optimal workshop and the corresponding routes that minimize $T$, we assume that:
1) The workshop that schedules the tow truck also conducts the maintenance;
2) Vehicle speeds on different roads are different;
3) Speeds of a loaded tow truck and an unloaded one are different;
4) Different workshops spend different time on scheduling a tow truck and performing the maintenance.

In the following subsections, we propose two methods to solve this problem.

### B. Method 1- *Mixed Integer Programming (MIP)*

An optimal route planning problem can be modeled using the network flow framework which turns out to be a MIP model. The MIP model can be solved directly using available optimization solvers such as GUROBI. The GUROBI solver uses the branch-and-bound algorithm to find the optimal solution of the MIP model.

We consider a set of workshop nodes represented by $\mathcal{N}^w = \{1, ..., N_1\}$, a set of traffic nodes represented by $\mathcal{N}^t = \{N_1+1, ..., N_2\}$, where node $N_1+1$ is the warehouse, node $N_2$ is the customer, and the rest of the nodes are interchanges. We consider a set of links $\mathcal{L}^t = \{1, ..., L\}$ connecting the nodes in $\mathcal{N}^w \cup \mathcal{N}^t$. The breakdown site is considered as a new node $N_2 + 1$, which introduces two new links $L+1$ and $L+2$. The road network is represented as a graph $G(\mathcal{N}, \mathcal{L}, \mathcal{A})$, in which $\mathcal{N}$ is the node set with $\mathcal{N} = \mathcal{N}^w \cup \mathcal{N}^t \cup \{N_2 + 1\}$, and $\mathcal{L} = \mathcal{L}^t \cup \{L+1, L+2\}$ is the link set. The links in $\mathcal{L}$ connect the nodes in $\mathcal{N}$.

To avoid the curse of dimensionality, the adjacency matrix $\mathcal{A}$ is built by representing the network topology with nodes and links (instead of only nodes). Similar methods have been used in [15], [16] and proved to be effective in improving the computational efficiency. For $n \in \mathcal{N}$, $j \in \mathcal{L}$, $A_{n,j} \in \{0, 1\}$, with $A_{n,j} = 0$ indicating that node $n$ and link $j$ are unconnected, and $A_{n,j} = 1$ indicating that node $n$ and link $j$ are connected.

We use the binary variable $x_i$, with $i \in \mathcal{N}^w$, $x_i \in \{0, 1\}$. $x_i = 0$ means that workshop $i$ is not selected, while $x_i = 0$ means that it is selected. We use the binary variable $y_{k,j}$, with $k \in \mathcal{R}$ and $j \in \mathcal{L}$, $y_{k,j} \in \{0, 1, -1\}$. $y_{k,j} = 0$ means that link $j$ is not selected in route $k$. y=1 means the link is selected and the driving direction is the same as the reference direction of the link. y=-1 means the link is selected and the driving direction is the opposite of the reference direction of the link.

The vehicle rerouting problem explained in Section III-A can be modelled as an MIP problem:

$$\text{Minimize} \underbrace{\sum_{i \in \mathcal{N}^w} x_i (T_i^{st} + T_i^m)}_{(2.1)} + \underbrace{\sum_{k \in \mathcal{R}} \sum_{j \in \mathcal{L}} (y_{k,j})^2 \frac{D_j}{V_{k,j}}}_{(2.2)} \quad (2)$$

subject to:

$$x_i \in \{0, 1\}, \quad \forall i \in \mathcal{N}^w \quad (3)$$

$$\sum_{i \in \mathcal{N}^w} x_i = 1, \quad (4)$$

$$y_{k,j} \in \{0, 1, -1\}, \quad \forall k \in \mathcal{R} \quad \forall j \in \mathcal{L} \quad (5)$$

$$\sum_{j \in \mathcal{L}} A_{N_2+1,j} \cdot y_{k,j} = -1, \quad \forall k \in \{1\} \quad (6)$$

$$y_{k,L+2} = 0, \quad \forall k \in \{1\} \quad (7)$$

$$\sum_{j \in \mathcal{L}} A_{N_2+1,j} \cdot y_{k,j} = 1, \quad \forall k \in \{2\} \quad (8)$$

$$y_{k,L+1} = 0, \quad \forall k \in \{2\} \quad (9)$$

$$\sum_{j \in \mathcal{L}} A_{N_2,j} \cdot y_{k,j} = -1, \quad \forall k \in \{3\} \quad (10)$$

$$y_{k,L+1} = y_{k,L+2}, \quad \forall k \in \{3\} \quad (11)$$

$$\sum_{j \in \mathcal{L}} A_{i,j} \cdot y_{k,j} = x_i, \quad \forall i \in \mathcal{N}^w, \quad \forall k \in \{1, 3\} \quad (12)$$

$$\sum_{j \in \mathcal{L}} A_{i,j} \cdot y_{k,j} = -x_i, \quad \forall i \in \mathcal{N}^w, \quad \forall k \in \{2\} \quad (13)$$

$$\sum_{j \in \mathcal{L}} A_{n,j} \cdot y_{k,j} = 0, \quad \forall n \in \mathcal{N}^t, \quad \forall k \in \{1, 2\} \quad (14)$$

$$\sum_{j \in \mathcal{L}} A_{n,j} \cdot y_{k,j} = 0, \quad \forall n \in \mathcal{N}^t \setminus \{N_2\}, \quad \forall k \in \{3\} \quad (15)$$

The objective function, Eq.(2) consists of two parts, where (2.1) refers to the time of scheduling the tow truck and conducting maintenance by workshop located at node $i$, and

(2.2) refers to the driving time on all the selected links in all the three routes.

$T_i^{st}$ and $T_i^m$ refer to the time of scheduling a tow truck and the time of maintenance by workshop $i$ respectively. $D_j$ is the length of link $j$ and $V_{k,j}$ is the average driving speed on link $j$ in route $k$.

Constraint (3) means that $x_i$ is a binary variable. Constraint (4) means that only one workshop is selected in the solution. Constraint (5) means the value set of variable $y_{k,j}$. Constraint (6) and constraint (7) mean that the breakdown site is the ending node of route 1. Constraint (8) and constraint (9) mean that the breakdown site is the starting node of route 2. Constraint (10) means that the final destination of the vehicle is the customer node. Constraint (11) means that after the maintenance, there is no need to distinguish link $L+1$ and $L+2$. Constraint (12) means that the origin of the vehicle in route 2 is the workshop node. Constraint (13) means that the destination of the vehicle is the workshop node. Constraints (14) and (15) mean that if the vehicle enters into an interchange node, it must leave the interchange node by using the connected links to that interchange node.

### C. Method 2 - Two-stage optimization based on Dijkstra's algorithm

As stated in Section III-A, the total time to deliver $T$ is relevant to the selected workshop. Therefore, Eq.(1) can be expressed as

$$T_i = T_i^{st} + T_i^m + \sum_{k \in \mathcal{R}} T_{k,i} \quad (16)$$

where $T_{k,i}$ is the driving time of route $k$ if workshop $i$ is selected.

To select a workshop so that the time to deliver is minimal, we can solve the optimization problem

$$\arg\min_{i \in \mathcal{N}^w} \quad T_i^{st} + T_i^m + \sum_{k \in \mathcal{R}} T_{k,i} \quad (17)$$

For a specific workshop and a specific route, the driving time of the route can be no less than the driving time of taking the shortest path, i.e.

$$T_{k,i} \geq T_{k,i}^{min} \quad \forall k \in \mathcal{R}, \quad \forall i \in \mathcal{N}^w \quad (18)$$

where $T_{k,i}^{min}$ is the driving time of the shortest path of route $k$ by selecting workshop $i$.

As a result, Eq.(17) is equivalent to

$$\arg\min_{i \in \mathcal{N}^w} \quad T_i^{st} + T_i^m + \sum_{k \in \mathcal{R}} T_{k,i}^{min} \quad (19)$$

which can be solved in two steps:
1) Obtain $T_{k,i}^{min}$ by searching the shortest path of each route given the workshop.

Consider the same node set $\mathcal{N}$ as in *Method 1*. Build two node sets $\mathcal{N}^{p1}$ and $\mathcal{N}^{p2}$, all the elements of which belongs to $\mathcal{N}$. The length of $\mathcal{N}^{p1}$ and $\mathcal{N}^{p2}$ are the same. For the $z_{th}$ elements in $\mathcal{N}^{p1}$ and $\mathcal{N}^{p2}$, $\mathcal{N}^{p1}[z]$ and $\mathcal{N}^{p2}[z]$ are connected, which is referred to as a node pair.

The road network is represented with graph $G'(\mathcal{N}^{p1}, \mathcal{N}^{p1}, \mathcal{W})$. $\mathcal{W}$ is the weight matrix, the element of which is the weight of the link connected by a node pair. The weight of a link represents the time to drive across this link, which is obtained with the length of this link and the average speed. In this paper, we assign each link a fixed value as the average driving speed.

Given the road network $G'$, the search of the shortest path of each route $k \in \mathcal{R}$ given workshop $i \in \mathcal{N}^w$ is implemented by Dijkstra's algorithm, which is a fundamental shortest path search algorithm.

2) Solve Eq.(19) and obtain the optimal workshop so that the total time to deliver is minimal.

## IV. NUMERICAL STUDY

We implement the two methods presented in Section III in MATLAB. The first method MIP uses the YALMIP optimization package and the GUROBI solver is deployed to solve the MIP model. The second method uses the Graph and Network Algorithms package to search for the shortest paths. The numerical performances of the two methods are demonstrated in this section by using three case studies, one base case with highway transport network, one modified case with highway transport network, and one case with urban city transport network. A personal computer with Intel Core-i7 3GHz CPU running on the Window 10 operating system is deployed to run the MATLAB codes.

### A. Base case with highway transport network

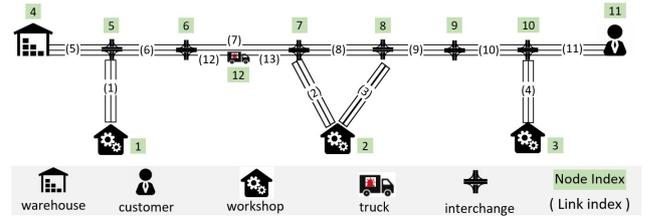

Figure 2. Transport scenario with a highway road network.

A base case with a highway road network is shown in Figure 2. In this case, we have three workshops indexed by the set $\mathcal{N}^w = \{1, 2, 3\}$, eight traffic nodes indexed by the set $\mathcal{N}^t = \{4, 5, 6, ...11\}$, a node of the breakdown site indexed by $\{12\}$, thirteen links indexed by the set $\mathcal{L} = \{1, 2, 3, ...13\}$.

The node-to-link adjacency matrix $\mathcal{A}$ in *Method 1* is shown in Table I. The reference direction of each link is set as from the positive element to the negative element in $\mathcal{A}$. For example, the reference direction of link-1 is from node-1 to node-5. This means if the solution of $y_{k,j}$ is equal to 1, the driving direction is the same as the reference direction of the corresponding link. Otherwise, if the solution of $y_{k,j}$ is equal to $-1$, it means the driving direction is the reverse of the reference direction of the corresponding link. In this set-up, there are three binary variables $x_i$ and thirty-nine integer variables $y_{k,j}$ to be solved from the MIP model.

Table I
TRANSPORTATION NETWORK ADJACENCY MATRIX $\mathcal{A}$ BY *Method 1*

| $i,j$ | $j=1$ | $j=2$ | $j=3$ | $j=4$ | $j=5$ | $j=6$ | $j=7$ | $j=8$ | $j=9$ | $j=10$ | $j=11$ | $j=12$ | $j=13$ |
|---|---|---|---|---|---|---|---|---|---|---|---|---|---|
| $i=1$ | 1 | 0 | 0 | 0 | 0 | 0 | 0 | 0 | 0 | 0 | 0 | 0 | 0 |
| $i=2$ | 0 | 1 | 1 | 0 | 0 | 0 | 0 | 0 | 0 | 0 | 0 | 0 | 0 |
| $i=3$ | 0 | 0 | 0 | 1 | 0 | 0 | 0 | 0 | 0 | 0 | 0 | 0 | 0 |
| $i=4$ | 0 | 0 | 0 | 0 | 1 | 0 | 0 | 0 | 0 | 0 | 0 | 0 | 0 |
| $i=5$ | -1 | 0 | 0 | 0 | -1 | 1 | 0 | 0 | 0 | 0 | 0 | 0 | 0 |
| $i=6$ | 0 | 0 | 0 | 0 | 0 | -1 | 1 | 0 | 0 | 0 | 0 | 1 | 0 |
| $i=7$ | 0 | -1 | 0 | 0 | 0 | 0 | -1 | 1 | 0 | 0 | 0 | 0 | -1 |
| $i=8$ | 0 | 0 | -1 | 0 | 0 | 0 | 0 | -1 | 1 | 0 | 0 | 0 | 0 |
| $i=9$ | 0 | 0 | 0 | 0 | 0 | 0 | 0 | 0 | -1 | 1 | 0 | 0 | 0 |
| $i=10$ | 0 | 0 | 0 | -1 | 0 | 0 | 0 | 0 | 0 | -1 | 1 | 0 | 0 |
| $i=11$ | 0 | 0 | 0 | 0 | 0 | 0 | 0 | 0 | 0 | 0 | -1 | 0 | 0 |
| $i=12$ | 0 | 0 | 0 | 0 | 0 | 0 | 0 | 0 | 0 | 0 | 0 | -1 | 1 |

Table II
LINKED NODE PAIRS BY *Method 2*.

| $\mathcal{N}^{p1}$ | 1 | 2 | 2 | 3 | 4 | 5 | 5 | 5 | 6 | 6 | 6 | 7 | 7 | 7 | 8 | 8 | 8 | 9 | 9 | 10 | 10 | 10 | 11 | 12 |
|---|---|---|---|---|---|---|---|---|---|---|---|---|---|---|---|---|---|---|---|---|---|---|---|---|
| $\mathcal{N}^{p2}$ | 5 | 7 | 8 | 10 | 5 | 1 | 4 | 6 | 5 | 7 | 12 | 2 | 6 | 8 | 2 | 7 | 9 | 8 | 10 | 3 | 9 | 11 | 10 | 7 |

The node sets $\mathcal{N}^{p1}$ and $\mathcal{N}^{p2}$ in *Method 2* are shown in Table II. One node pair $(i,n)$ represents a directional link from node-$i$ to node-$n$. In total, there are 24 directional links in the road network shown in Figure 2.

For ease of illustration, we set the parameters of the tow-scheduling time as $T_i^{st} = 10$ min, $\forall i \in \{1,2,3\}$, the maintenance time as $T_i^m = 100$ min, $\forall i \in \{1,2,3\}$, the link length as $D_j = 10$ km, $\forall j \in \{1,2,3,...11\}$ and $D_j = 5$ km, $\forall j \in \{12,13\}$, the average driving speed as $V_{k,j} = 60$ km/h, $\forall k \in \{1,2,3\}, \forall j \in \{1,2,3...13\}$.

*1) Method 1 - MIP:* Solving the MIP model in MATLAB, the solutions of the decision variables $x_i$ and $y_{k,j}$ are listed in Table III and Table IV. In the last row of Table IV, we also list the links of the optimal paths for all three routes.

In this case, the objective solution i.e the minimum of the total time equal to 190 min which includes 10 min to schedule the tow truck, 100 min of maintenance work, and 80 min of driving time. The computational CPU time in solving this MIP model is 0.14 seconds which shows a high efficiency of the MIP model.

For the decision variable $x_i$, as shown in Table III, the GUROBI solver selects the second workshop located at node-2 for the maintenance work. Intuitively, this is optimal since this workshop is the closest one to the breakdown site and it takes the least time to drive there.

For the solutions of the decision variable $y_{k,j}$, the results are shown in Table III. It is worth mentioning that the driving direction of the highway for the link-12, the link-13, and the link-7 is fixed. This means it is not permitted to go directly from node-7 to node-12 across link-13 (from the workshop to the breakdown site). The GUROBI solver finds the shortest path through link-7 and link-12 for route 1, the shortest path through link-13 and link-12 for route 2, and the shortest path through link-3, link-9, link-10, and link-11 for route 3. All these solutions are optimal in terms of minimizing the maintenance time and total driving time for all three routes.

Table III
SOLUTION OF $x_i$ OF THE BASE CASE BY *Method 1*

| $i$ | 1 | 2 | 3 |
|---|---|---|---|
| $x_i$ | 0 | 1 | 0 |

Table IV
SOLUTION OF $y_{k,j}$ OF THE BASE CASE BY *Method 1*

| $k,j$ | $k=1$ | $k=2$ | $k=3$ |
|---|---|---|---|
| $j=1$ | 0 | 0 | 0 |
| $j=2$ | 1 | -1 | 0 |
| $j=3$ | 0 | 0 | 1 |
| $j=4$ | 0 | 0 | 0 |
| $j=5$ | 0 | 0 | 0 |
| $j=6$ | 0 | 0 | 0 |
| $j=7$ | -1 | 0 | 0 |
| $j=8$ | 0 | 0 | 0 |
| $j=9$ | 0 | 0 | 1 |
| $j=10$ | 0 | 0 | 1 |
| $j=11$ | 0 | 0 | 1 |
| $j=12$ | 1 | 0 | 0 |
| $j=13$ | 0 | -1 | 0 |
| Path Links | 2, 7, 12 | 13, 2 | 3, 9, 10, 11 |

*2) Method 2 - Two-stage optimization:* Using Dijkstra's algorithm, given the selected workshop, the shortest path of each of the three routes and the corresponding driving time it takes are listed in Table V. For any selected workshop, the minimal total time to deliver is listed in Table VI. It is shown in Table VI that by selecting workshop at node-

2, the minimal total time to deliver is minimal among the three workshops. Therefore, the workshop located at node-2 is optimal for the maintenance work. The corresponding optimal paths can be retrieved in Table V.

The results are the same as with *Method 1*. A total time of 0.017 seconds is taken to run the relevant codes in Matlab, which is significantly less than that of *Method 1*.

Table V
SHORTEST PATHS OF THE THREE ROUTES OF THE BASE CASE OF *Method 2*

| $k$ | starting node | ending node | shortest path nodes | $T_{k,i}^{min}$ |
|---|---|---|---|---|
| 1 | 1 | 12 | [1,5,6,12] | 25 |
| 1 | 2 | 12 | [12,7,6,5,1] | 35 |
| 1 | 3 | 12 | [1,5,6,7,8,9,10,11] | 70 |
| 2 | 12 | 1 | [2,7,6,12] | 25 |
| 2 | 12 | 2 | [12,7,2] | 15 |
| 2 | 12 | 3 | [2,8,9,10,11] | 40 |
| 3 | 1 | 11 | [3,10,9,8,7,6,12] | 55 |
| 3 | 2 | 11 | [12,7,8,9,10,3] | 45 |
| 3 | 3 | 11 | [3,10,11] | 20 |

Table VI
MINIMAL TOTAL TIME TO DELIVER GIVEN THE WORKSHOP OF *Method 2*

| workshop $i$ | 1 | 2 | 3 |
|---|---|---|---|
| $T_i^{min}$ (minutes) | 240 | 190 | 230 |

### B. Modified case with highway transport network

With the transport scenario in Figure 2, we modify the parameters of the tow truck scheduling time as $T_1^{st} = 10$ min, $T_2^{st} = 20$ min, $T_3^{st} = 30$ min, the maintenance time as $T_1^m = 100$ min, $T_2^m = 200$ min, $T_3^m = 300$ min, the link length as $D_1 = 10$ km, $D_2 = 50$ km, $D_3 = 60$ km, $D_4 = 30$ km, $D_5 = 20$ km, $D_6 = 15$ km, $D_7 = 120$ km, $D_8 = 40$ km, $D_9 = 30$ km, $D_{10} = 60$ km, $D_{11} = 30$ km, $D_{12} = 30$ km, $D_{13} = 90$ km, the average driving speed as $V_{1,j} = 60$ km/h, $V_{2,j} = 30$ km/h, $V_{3,j} = 60$ km/h, $\forall j \in \{1,2,3,4\}$ (to model the non-urban roads), $V_{1,j} = 100$ km/h, $V_{2,j} = 80$ km/h, $V_{3,j} = 100$ km/h, $\forall j \in \{5,6,7...13\}$ (to model the highway).

*1) Method 1 - MIP:* In this set-up, the optimal solutions of the decision variables $x_i$ and $y_{k,j}$ using the MIP Method are listed in Table VII and Table VIII. In the last row of Table VIII, we also list the links of the optimal paths for all the three routes. The objective solution i.e the minimum of the total time equal to 522.75 min which includes 10 min to schedule the tow truck, 110 min of maintenance work, and 512.75 min of driving time. The computational CPU time in solving this MIP model is 0.17 seconds which again shows a high efficiency of our MIP model.

For the decision variable $x_i$ as shown in Table VII, the GUROBI solver selects the first workshop located at node-1 for the maintenance work. This is optimal since this workshop takes the shortest scheduling time and maintenance time even considering the driving distances in this case.

The solutions of the decision variable $y_{k,j}$ are listed in Table VII. The GUROBI solver finds the shortest path through link-1, link-6 and link-12 for route 1, the shortest path through link-13, link-7, link-6 and link-1 for route 2, and the shortest path is through link-1, link-6, link-12, link-13, link-8, link-9, link-10 and link-11 for route 3. All these solutions are optimal in terms of minimizing the maintenance time and total driving time jointly considering all three routes.

Table VII
SOLUTION OF $x_i$ OF THE MODIFIED CASE OF *Method 1*

| $i$ | 1 | 2 | 3 |
|---|---|---|---|
| $x_i$ | 1 | 0 | 0 |

Table VIII
SOLUTION OF $y_{k,j}$ OF THE MODIFIED CASE OF *Method 1*

| $k,j$ | $k=1$ | $k=2$ | $k=3$ |
|---|---|---|---|
| $j=1$ | 1 | -1 | 1 |
| $j=2$ | 0 | 0 | 0 |
| $j=3$ | 0 | 0 | 0 |
| $j=4$ | 0 | 0 | 0 |
| $j=5$ | 0 | 0 | 0 |
| $j=6$ | 1 | -1 | 1 |
| $j=7$ | 0 | -1 | 0 |
| $j=8$ | 0 | 0 | 1 |
| $j=9$ | 0 | 0 | 1 |
| $j=10$ | 0 | 0 | 1 |
| $j=11$ | 0 | 0 | 1 |
| $j=12$ | 1 | 0 | 1 |
| $j=13$ | 0 | -1 | 1 |
| Path Links | 1, 6, 12 | 13, 7, 6, 1 | 1, 6, 12, 13, 8, 9, 10, 11 |

*2) Method 2 - Two-stage optimization:* Given any workshop, the shortest path of each of the three routes happens to be the same as with the base case, which can be retrieved from Table V. The minimal total time to deliver is listed in Table IX. As shown in the table, by selecting workshop with node-1, the minimal total time to deliver $T_i^{min}$ is minimal among the three workshops. Therefore, the workshop located at node-1 is optimal for the maintenance work. The corresponding minimal total time to deliver is 522.75 min.

The results are the same as with *Method 1*. A total time of 0.017 seconds is taken to run the relevant codes in Matlab, which is significantly less than that of *Method 1*.

Table IX
MINIMAL TOTAL TIME TO DELIVER GIVEN THE WORKSHOP OF *Method 2*

| workshop $i$ | 1 | 2 | 3 |
|---|---|---|---|
| $T_i^{min}$ (minutes) | 522.75 | 659.5 | 801 |

### C. Case with urban transport network

A use case with an urban transport network is shown in Figure 3. Note the link (road) lengths in Figure 3 are not proportionally scaled. They are only to indicate the topology of the city road network. In this example, there are six workshops, twenty-six nodes and forty-two road links. The parameters of the tow-scheduling times are $T_1^{st} = 10$ min, $T_2^{st} = 20$ min, $T_3^{st} = 30$ min, $T_1^{st} = 40$ min, $T_2^{st} = 50$ min, $T_3^{st} = 60$ min, the maintenance times are $T_1^m = 600$ min, $T_2^m = 500$ min, $T_3^m = 400$ min, $T_1^m = 300$ min, $T_2^m = 200$ min, $T_3^m = 100$ min, the link lengths are listed in Table X, the average driving speed are $V_{1,j} = 30$ km/h, $V_{2,j} = 20$ km/h, $V_{3,j} = 30$ km/h, $\forall j \in \{1, 2, 3, ...42\}$ (to model the urban roads).

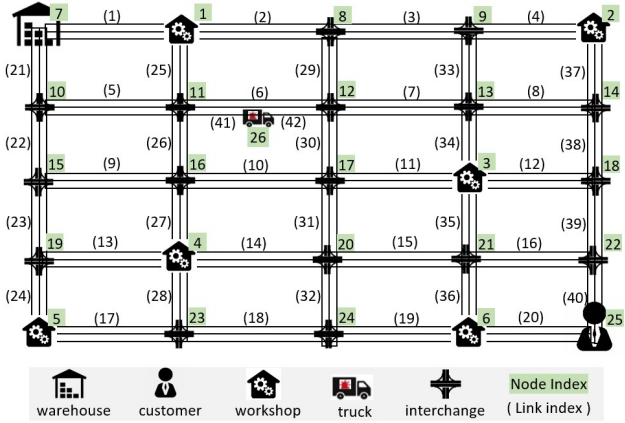

Figure 3. Transport scenario with an urban road network.

Table X
LINK LENGTH OF THE LARGE-SCALE EXAMPLE

| $j$ | $D_j$ | $j$ | $D_j$ | $j$ | $D_j$ | $j$ | $D_j$ | $j$ | $D_j$ |
|---|---|---|---|---|---|---|---|---|---|
| 1 | 10 | 10 | 60 | 19 | 15 | 28 | 50 | 37 | 30 |
| 2 | 50 | 11 | 30 | 20 | 120 | 29 | 60 | 38 | 30 |
| 3 | 60 | 12 | 30 | 21 | 40 | 30 | 30 | 39 | 90 |
| 4 | 30 | 13 | 90 | 22 | 30 | 31 | 20 | 40 | 100 |
| 5 | 20 | 14 | 10 | 23 | 60 | 32 | 15 | 41 | 60 |
| 6 | 150 | 15 | 50 | 24 | 30 | 33 | 120 | 42 | 90 |
| 7 | 120 | 16 | 60 | 25 | 30 | 34 | 40 | | |
| 8 | 40 | 17 | 30 | 26 | 90 | 35 | 30 | | |
| 9 | 30 | 18 | 20 | 27 | 10 | 36 | 60 | | |

*1) Method 1 - MIP:* The optimal solutions of the decision variables $x_i$ and $y_{k,j}$ using the MIP Method are listed in Table XI and Table XII. Note that in Table XII, only the links $j$ that correspond to at least one non-zero $k$ are listed to make the table more concise. In the last row of Table XII, we also list the links of the optimal paths for all three routes. The objective solution i.e the minimum of the total time equal to 1290 min which includes 60 min to schedule the tow truck, 100 min of maintenance work, and 1130 min of driving time. The computational CPU time in solving this MIP model is 0.29 seconds which again shows a high efficiency of our MIP model.

For the decision variable $x_i$, as shown in Table XI, the GUROBI solver selects the sixth workshop located at node-6 for the maintenance work. This is optimal since this work-shop takes the shortest scheduling time and maintenance time considering the driving distances in this case.

The solutions of the decision variable $y_{k,j}$ are listed in Table XI. The GUROBI solver finds the shortest path through link-19, link-32, link-14, link-27, link-9, link-22, link-5 and link-41 for route 1, the shortest path through link-42, link-30, link-31, link-32 and link-19 for route 2 and the shortest path through link-20 for route 3. All these solutions are optimal in terms of minimizing the maintenance time and total driving time jointly considering all three routes.

Table XI
SOLUTION OF $x_i$ OF THE LARGE-SCALE EXAMPLE OF *Method 1*

| $i$ | 1 | 2 | 3 | 4 | 5 | 6 |
|---|---|---|---|---|---|---|
| $x_i$ | 0 | 0 | 0 | 0 | 0 | 1 |

Table XII
SOLUTION OF $y_{k,j}$ OF THE LARGE-SCALE EXAMPLE OF *Method 1*

| $j$ | $k=1$ | $k=2$ | $k=3$ |
|---|---|---|---|
| $j=5$ | 1 | 0 | 0 |
| $j=9$ | -1 | 0 | 0 |
| $j=14$ | -1 | 0 | 0 |
| $j=19$ | -1 | 1 | 0 |
| $j=22$ | -1 | 0 | 0 |
| $j=27$ | -1 | 0 | 0 |
| $j=30$ | 0 | 1 | 0 |
| $j=31$ | 0 | 1 | 0 |
| $j=32$ | -1 | 1 | 0 |
| $j=41$ | 1 | 0 | 0 |
| $j=42$ | 0 | 1 | 0 |
| Path Links | 19, 32, 14, 27, 9, 22, 5, 41 | 42, 30, 31, 32, 19 | 20 |

*2) Method 2 - Two-stage optimization:* For any selected workshop, the minimal total time to deliver is presented in Table XIII. As shown in the table that by selecting workshop with node-6, the minimal total time to deliver $T_6^{min}$ is 1290 min, which is minimal among the three workshops. Therefore, this workshop is optimal for maintenance work. The corresponding node sets representing the three routes are $\{6, 24, 20, 4, 16, 15, 10, 11, 26\}$, $\{26, 12, 17, 20, 24, 6\}$ and $\{6, 25\}$, which is the same as with *Method 1*. It takes a total time of 0.021 seconds to run the relevant codes in Matlab, which is significantly less than that of *Method 1*.

Table XIII
MINIMAL TOTAL TIME TO DELIVER GIVEN THE WORKSHOP OF *Method 2*

| workshop $i$ | 1 | 2 | 3 | 4 | 5 | 6 |
|---|---|---|---|---|---|---|
| $T_i^{min}$ (minutes) | 1950 | 2200 | 1680 | 1410 | 1635 | 1290 |

*D. Discussion*

It is shown that the two proposed methods are both feasible for the formulated decision-making problem and they give the same solution for the different cases. However, the computational efficiency of the two-stage optimization method far outperforms that of the MIP method in all the test cases. The reason could be that 1) by splitting the optimization into two stages, the solution space is largely reduced; 2) the road network modeling of the MIP method is more complex and redundant than that of the other method by considering the links that are not physically connected.

However, it is worth mentioning that compared to the MIP method, the feasibility of the two-stage optimization method relies on the assumption that 'The workshop that schedules the tow truck also conducts the maintenance'. If this assumption is released, the optimization problem can not be split into two stages and solved separately. From this perspective, the MIP method has a better extendability than the other one when the complexity of the problem increases.

## V. CONCLUSION

In this paper, we identify the challenges of real-time decision-making of autonomous vehicles under faults from the maintenance management perspective. A decision-making problem of a vehicle breakdown during a transport mission is formulated as a vehicle rerouting problem considering maintenance management. We propose two methods to solve the problem, the MIP method and the two-step optimization based on Dijkstra's algorithm. The former method builds an explicit mathematical optimization model featuring reducing the number of decision variables by indexing the routes by links instead of by the nodes. The latter method decomposes the optimization problem into two steps and adopts Dijkstra's algorithm to search the shortest path for one step. Three numerical case studies show that both methods converge to the same route solution. The computational efficiency of the two-step optimization method outperforms that of the MIP method, while its extendability is inferior to the MIP method for more complex objective functions or constraints.

In future research, the decision-making problem will be expanded to a broader extent regarding complexity and uncertainty. Specifically, uncertain fault conditions of the vehicle and more maintenance resources such as spare vehicles will be considered. To solve more complex problems, decision-making methods that consider both computationally efficiency and extendability will be developed.

## CREDIT (CONTRIBUTOR ROLES TAXONOMY) AUTHORSHIP CONTRIBUTION STATEMENT

**Xin Tao**: Methodology, Software, Investigation, Writing - Original Draft, Writing - Review & Editing. **Zhao Yuan**: Methodology, Software, Investigation, Writing - Original Draft, Writing - Review & Editing.